\documentstyle[12pt]{article}

\def\pd{\partial}
\def\a{\alpha}
\def\b{\beta}
\def\dl{\delta}
\def\s{\sigma}

\def\lam{\lambda}

\def\bg{{\bar g}}
\def\hg{{\hat g}}

\def\hnabla{{\hat \nabla}}

\def\bR{{\bar R}}

\def\bG{{\bar G}}

\def\bC{{\bar C}}
\def\bDelta{{\bar \Delta}}
\def\bBox{\stackrel{-}{\Box}}

\def\gm{\gamma}

\def\dota{\stackrel{.}{a}}

\def\dotH{\stackrel{.}{H}}
\def\ddotH{\stackrel{..}{H}}
\def\dddotH{\stackrel{...}{H}}
\def\sq{\sqrt}
\def\e{\hbox{\large \it e}}

\def\fr{\frac}

\def\arr{\rightarrow}

\def\bb{\begin{equation}}
\def\ee{\end{equation}}
\def\bba{\begin{eqnarray}}
\def\eea{\end{eqnarray}}

\begin{document}

\begin{titlepage}

\begin{tabbing}
   qqqqqqqqqqqqqqqqqqqqqqqqqqqqqqqqqqqqqqqqqqqqqq 
   \= qqqqqqqqqqqqq  \kill 
         \>  {\sc KEK-TH-738}    \\
         \>       hep-th/0101100 \\
         \>  {\sc January, 2001} 
\end{tabbing}
\vspace{1.5cm}

\begin{center}
{\Large {\bf A Dynamical Solution of 
Stable Starobinsky-Type Inflationary Model \\
in Quantum Geometry}}
\end{center}

\vspace{1.5cm}

\centering{\sc Ken-ji Hamada\footnote{E-mail address : 
hamada@post.kek.jp} }

\vspace{1cm}

\begin{center}
{\it Institute of Particle and Nuclear Studies, \break 
High Energy Accelerator Research Organization (KEK),} \\ 
{\it Tsukuba, Ibaraki 305-0801, Japan}
\end{center} 

\vspace{1.5cm}

\begin{abstract} 
Quantum geometry gives a regularization scheme-independent 
effective action, whoes equation of motion for the conformal mode  
has a stable de Sitter solution 
at the high-energy region where the coupling of the 
self-interactions of the traceless mode can be neglected 
because of the asymptotic freedom.  
However, the dynamics of the traceless mode suggests that 
inflation ends at the low-energy region. 


\end{abstract}
\end{titlepage}

Inflationary scenarios have been studied as the most simple idea 
to solve both the flatness and the horizon problems 
naturally~\cite{gu}--\cite{fps}.  
However, standard inflationary models~\cite{gu,li} 
have demerits, such as we must 
introduce an unsatisfactory quantity like inflaton and 
need a fine-tuning to get a quantitatively reasonable scenarios.  
On the other hand, there is an alternative model by Starobinsky~\cite{st}, 
who propose a scenario that inflation is driven dynamically 
due to one-loop corrections by quantized matter fields. 
He originally consider unstable de Sitter phase  
by adjusting the regularization sheme-dependent term properly 
because inflation has to end at the low-energy phase. 
However, even in this scenario,  a fine-tuning of the 
scheme-dependent parameter is necessary to obtain sufficiently long 
lifetime~\cite{vi,hhr}. 

The inflation models mentioned above are 
based on semi-classical arguments. 
On the other hand, it is believed that quantum geometry solve 
many problems in quantum cosmology~\cite{t,am}.  
In this letter we show that quantum geometry can really solve 
such fine-tuning problems. 
The idea is based on the recent developments~\cite{h00,h99}  
in which we show that the effective action of 4D quantum geometry has 
a regularization scheme-independent form  
and may be uniquely determined according to conformal matter contents.  
The equation of motion of this effective 
action has a stable de Sitter solution. 
A mechanism to end the inflationary phase is given by the 
dynamics of the traceless. 
 
Let us first summarize the results given in ref.~\cite{h00}.
The form of effective action of 4D quantum geometry 
is strongly constrained by diffeomorphism invariance/background-metric 
independence~\cite{h99,h00}. 
It is because a general coordinate transformation in quantum geometry 
is related to a conformal change of the background metric 
so that the integrability condition~\cite{wz,bcr} 
of conformal anomalies~\cite{cd,ddi,du} 
plays an important role to determine the  effective action. 
Really, as discussed in~\cite{h99,h00,r,ft1}, 
the integrability condition of conformal anomalies 
not only restricts matter fields to be 
conformally invariant, but also determines many indefiniteness in 
the gravity sector. 
To preserve diffeomorphism invariance at the quantum level, 
we must add the Wess-Zumino action to an invariant action 
and quantize such a combined action as a tree action  
in a self-consistent mannar~\cite{p,amm,hs,h99,h00}.   
The scheme-dependent terms which appear in  
loop effects of the combined theory 
cancel with the corresponding terms in the Wess-Zumino 
action, and then diffeomorphism invariance is realized~\cite{h00}.   
       
We consider a perturbation theory in which the traceless mode 
is expanded by the dimensionless coupling, $t$, as 
$\bg_{\mu\nu} =(\hg\e^{th})_{\mu\nu}$~\cite{kkn,hs,h99,h00}. 
Here, the metric is decomposed 
as $g_{\mu\nu}=\e^{2\phi}\bg_{\mu\nu}$. On the other hand, 
the conformal mode, $\phi$, is evaluated exactly, because we treat  
conformal anomalies.  
Thus we obtain a regularization scheme-independent form 
of 4D quantum geometry effective action 
(in Lorentzian signature)~\cite{h00},\footnote{
The scheme-dependent term has a non-invariant form, $\bR^2$, which 
have to cancel out from the requirement of diffeomorphism invariance. 
While an invariant $R^2$ term might appear at least from 
order $t^4_r$~\cite{h99}.
}
\bb
    \Gamma  
    = \fr{1}{(4\pi)^2}\int d^4 x \sq{-g} \biggl\{  
         \frac{f}{4}  C_{\mu\nu\lam\s} 
           \log\biggl( \fr{\Delta^C_4}{\mu^4} \biggr) C^{\mu\nu\lam\s}         
        -\frac{e}{8} {\cal G} \fr{1}{\Delta_4} {\cal G} \biggr\}
          + I_{LE} , 
\ee
where $I_{LE}$ represents lower derivative actions which include actions 
of conformally invariant matter fields and the Einstein-Hilbert action. 
$C_{\mu\nu\lam\s}$ is the Weyl tensor and 
$\Delta^C_4 = \Box^2 + \cdots$ 
is an appropriate conformally covariant operator for the Weyl tensor. 
The explicit form of $\Delta^C_4$ 
is unknown, but it is known that the conformal variation of this term 
produces the square of the Weyl tensor~\cite{ddi,ds,deser}.  
${\cal G}$ is defined by the following combination:   
\bb
    {\cal G} = G - \frac{2}{3}\Box R  , 
\ee
where $G$ is the Euler density defined by 
\bb
    G = R_{\mu\nu\lam\s}R^{\mu\nu\lam\s}-4R_{\mu\nu}R^{\mu\nu} 
           +R^2 . 
\ee
$\Delta_4$ is the conformally covariant 4th 
order operator~\cite{r}  
\bb
      \Delta_4 = \Box^2 
               + 2 R^{\mu\nu}\nabla_{\mu}\nabla_{\nu} 
                -\fr{2}{3}R \Box 
                + \fr{1}{3}(\nabla^{\mu}R)\nabla_{\mu} ,  
\ee
which satisfies $\Delta_4 = \e^{-4\phi}\bDelta_4 $ locally for a scalar. 

The coefficients $f$ and $e$ are scheme-independent. 
They are expanded by the renormalized 
coupling, $t_r$, as 
\bb 
   f=f_0 + f_1 t_r^2  + \cdots , \qquad   
   e=e_0 + e_1 t_r^2  +\cdots .   
\ee
Here, $f_0$ and $e_0$ have already been computed by one-loop diagrams 
as 
\bba
    && f_0 = -\fr{N_X}{120}-\fr{N_W}{40} -\fr{N_A}{10} 
           -\fr{199}{30}+\fr{1}{15} , 
                 \\ 
    && e_0 = \fr{N_X}{360} +\fr{11N_W}{720} +\fr{31N_A}{180} 
           +\fr{87}{20} -\fr{7}{90} , 
\eea 
where the first three contributions of each coefficient come from 
$N_X$ conformal scalar fields, $N_W$ Weyl fermions and $N_A$ gauge 
fields, respectively~\cite{du}. The fourth and the last ones  
come from the traceless mode~\cite{ft2} 
and the conformal mode~\cite{amm}, respectively.   
The coefficients $f_1$ and $e_1$ are contributions from 
not only two-loop diagrams, but also one-loop diagrams
of order $t_r^2$ which include vertices of the Wess-Zumino 
action~\cite{h99}.    

The beta function for the coupling, $t_r$, is given by 
$\b=\fr{f}{2} t_r^3$.  
Since $f_0$ is negative, 4D quantum geometry is asymptotically free. 
Here, note that, although background-metric independence includes 
an invariance under any confromal change of the background metric, 
the usual $\b$-function is not needed to vanish. This nature  
is because there is a conformal anomaly, or the Wess-Zumino 
action in 4 dimensions~\cite{h99,h00}. 

Consider the equation of motion for the conformal mode, 
\bb
    -\fr{f}{(4\pi)^2} \sq{-\bg} \bC^2 
    - \fr{e}{(4\pi)^2} \sq{-\bg}\biggl( 4 \bDelta_4 \phi 
            + \bG -\fr{2}{3} \bBox \bR \biggr) 
     + \fr{\dl}{\dl\phi} I_{EH}(g) =0 . 
             \label{eom}
\ee
Here, matter fields are conformally invariant so that 
there are no contributions to this equation of motion. 
The Einstein-Hilbert action has the following form:
\bb
    I_{EH} = 6m^2 \int d^4 x \sq{-\hg} \e^{\a\phi}
        \hnabla^{\lam}\phi \hnabla_{\lam}\phi + o(t_r^2) ,  
\ee 
where, $\a$ is defined through the anomalous dimension of $m$ by  
the relation $\a =2+\gm_m$. The lowest order of the anomalous 
dimension is computed by one-loop diagrams 
as $\gm_m =\fr{\a^2}{4e_0}$~\cite{am}. 
However, as for the minimal Standard Model with $N_X=4$, $N_W=45$ 
and $N_A=12$, for example, the anomalous dimension is sufficiently 
small as $\gm_m \sim 1/e_0 = 0.14$. 
Thus, in the following, we take the value of $\a$ approximately 
to be the classical one, $\a=2$. 

Since the dynamics of the traceless mode is asymptotically free, 
in the high-energy region, the coupling, $t_r$, becomes small and 
we can neglect it. 
Further, we neglect the space-coordinate dependence 
of the conformal mode. 
We also consider the conformally flat metric, because it is simple and 
also the recent experiments suggest the flat 
universe~\cite{flat}, as
\bb
        ds^2 = a(\eta)^2 \Bigl(-d\eta^2 +(dx^i)^2 \Bigr) .
                  \label{cfm}
\ee  
Now, equation (\ref{eom}) has a quite simple form,
\bb
      -\fr{4e}{(4\pi)^2} \pd_{\eta}^4 \log a  
      +12m^2 a^2 \Bigl( \pd_{\eta}^2 \log a 
         +(\pd_{\eta}\log a)^2 \Bigr) =0 . 
\ee 
Here, we introduce the proper time as $d\tau =a(\eta)d\eta$. 
and use the variable $H=\fr{\dota}{a}$, where a dot denotes 
a derivative with respect to the proper time. 
We then obtain the equation of motion for $H$ as 
\bb
    \dddotH +7 H\ddotH +4\dotH^2 +18H^2 \dotH +6H^4 
     - \fr{3M^2}{e} \Bigl( \dotH +2H^2 \Bigr) =0 ,    
          \label{eoh}
\ee
where $M=4\pi m$. This equation has a de Sitter solution
\bb
          \dotH=0 , \qquad H= H_0 , 
\ee
where $H_0 =\pm \fr{M}{\sq{e}}$. The plus sign gives an exponentially 
expanding solution, $a(\tau)=a_0 \e^{M\tau/\sq{e}}$. 

Next, we discuss the stability of this solution. 
Consider a small deviation from the de Sitter solution as 
$H=H_0 (1+\dl)$. Substituting this into equation (\ref{eoh}) and 
neglecting $o(\dl^2)$ terms, we obtain 
\bb
     \stackrel{...}{\dl} +7H_0 \stackrel{..}{\dl} 
     +15 H_0^2 \stackrel{.}{\dl} +12H_0^3 \dl =0 . 
\ee 
We here consider a solution of the type $\dl=\e^{\kappa\tau}$. 
We then obtain 
the equation, $\kappa^3 +7H_0 \kappa^2 +15H_0^2 \kappa +12H_0^3 =0$.  
This equation has three solutions
\bb
       -4H_0 , \qquad 
      \biggl( -\fr{3}{2} \pm i \fr{\sq{3}}{2} \biggr) H_0 . 
\ee
For $H_0 >0$, all solutions have $Re(\kappa) <0$. 
Thus, the de Sitter solution is stable. 

Inflation must end at the low-energy region. In our model 
a classical limit is given by $e \arr \infty$. 
The asymptotic freedom suggests that this limit will be realized 
at the low-energy region where the coupling, $t_r$, diverges.    
Large fluctuations of the gravitational fields freeze and 
small fluctuations ruled by the Einstein-Hilbert action survive
at the low-energy region. 
Thus, the lifetime of the inflationary phase is given by  
the order of the inverse of the scale, $\mu$. 

A stable inflationary scenario is also considered in~\cite{fps}, 
in which, however, there is no dynamical reason to end inflation  
because their argument is semi-classical.     

There are some problems in our scenario. One is that we set the 
cosmological constant vanishing as in the original paper by Starobinsky, 
because whether the de Sitter solution exists or not depends on the 
initial value of the cosmological constant.
The other is that we here restrict the solution of the equation of 
motion as in the form (\ref{cfm}) where the space-coordinate dependence 
of the conformal mode is neglected. 
In our scenario, vanishing of the scalar curvature is realized by the 
dynamics of the traceless mode. This is different from a mechanism 
discussed in~\cite{am}, in which the dynamics of the 4 derivative theory 
of the conformal mode gives a vanishing expectation 
value of the scalar curvature at the long distance.   
In any case, the dynamics of the gravitational fields 
would solve problems in inflationary cosmology.

\vspace{5mm}

\begin{flushleft}
{\bf Acknowledgements}
\end{flushleft}

  This work is supported in part by the Grant-in-Aid for 
Scientific Research from the Ministry of 
Education, Science and Culture of Japan.

\end{document}